\documentclass[fleqn,10pt]{wlscirep}
\usepackage[utf8]{inputenc}
\usepackage[T1]{fontenc}

\title{Hessian QM9: A quantum chemistry database of molecular Hessians in implicit solvents}

\author[1,*,$\dag$]{Nicholas J. Williams}
\author[2,$\dag$]{Lara Kabalan}
\author[2,$\dag$]{Ljiljana Stojanovic}
\author[2,$\dag$]{Viktor Z\'olyomi}
\author[1]{Edward O. Pyzer-Knapp}
\affil[1]{IBM Research Europe, United Kingdom}
\affil[2]{Hartree Centre, Science and Technology Facilities Council, Daresbury Laboratory, Daresbury, WA4 4AB, United Kingdom}

\affil[$\dag$]{these authors contributed equally to this work}

\affil[*]{corresponding author: Nicholas J. Williams (nicholas.williams@ibm.com)}

\begin{abstract}
A significant challenge in computational chemistry is developing approximations that accelerate \emph{ab initio} methods while preserving accuracy. Machine learning interatomic potentials (MLIPs) have emerged as a promising solution for constructing atomistic potentials that can be transferred across different molecular and crystalline systems. Most MLIPs are trained only on energies and forces in vacuum, while an improved description of the potential energy surface could be achieved by including the curvature of the potential energy surface. We present Hessian QM9, the first database of equilibrium configurations and numerical Hessian matrices, consisting of 41,645 molecules from the QM9 dataset at the $\omega$B97x/6-31G* level. Molecular Hessians were calculated in vacuum, as well as water, tetrahydrofuran, and toluene using an implicit solvation model. To demonstrate the utility of this dataset, we show that incorporating second derivatives of the potential energy surface into the loss function of a MLIP significantly improves the prediction of vibrational frequencies in all solvent environments, thus making this dataset extremely useful for studying organic molecules in realistic solvent environments for experimental characterization.
\end{abstract}
\begin{document}

\flushbottom
\maketitle

\thispagestyle{empty}

\section*{Background \& Summary}

Since the seminal work of Hohenberg and Kohn\cite{PhysRev.136.B864}, density functional theory (DFT) has emerged as the most popular first principles method in quantum chemistry. The rise of DFT can be attributed to its remarkable speed as compared to traditional \textit{ab initio} quantum chemistry methods. Despite its speed, and the advances in the computational efficiency of DFT including the development of linear scaling DFT\cite{skylaris2005introducing}, as well as advances in computing hardware, it is still infeasible to apply DFT to large system (>1000 atoms) and long time scales (>1 ns). To overcome this limitation, machine learning interatomic potentials (MLIPs) have been developed to perform atomistic simulations at a precision approaching that of DFT but at a cost closer to that of classical force fields. MLIP codes such as ANI\cite{gao2020ani, smith2020ani, devereux2020ani, smith2017ani}, CHGNet\cite{deng2023chgnet}, Nequip\cite{batzner20223nequip}, Allegro\cite{musaelian2023allegro}, MACE\cite{batatia2022mace} and  FieldSchNet\cite{gastegger2021machine} have gradually paved the way to an increasingly greater level of applicability across the chemical space under sufficiently diverse training data assumptions.

MLIP models heavily rely on high-quality and extensive reference data during their training process. The Chemical Space Project enumerated all feasible organic molecules up to 17 atoms leading to the creation of the GDB-17 databases, encompassing an impressive 166 billion molecules.\cite{reymond2015chemicalspace} QM9 is the gold standard benchmark dataset where various MLIPs have been, comprising of 133,885 ground-state geometries containing up to nine heavy atoms from the GDB-17 database (H, C, N, O, F).\cite{ramakrishnan2014qm9} QM9 provides ample information about each chemical species (energy, harmonic frequencies, dipole moments, polarizabilities, enthalpies, and free energies of atomization) where all properties were computed at the B3LYP/6–31G(2df,p) level of quantum chemistry. Despite the growing amount of open-sourced molecular data, few quantum chemical calculations are carried out in an implicit solvent environment.\cite{ward2021graph, gastegger2021machine} In this work, we have carried out geometry relaxations to the equilibrium ground state on the QM9 dataset in vacuum, water, tetrahydrofuran and toluene.

\begin{figure}[ht]
  \centering
  \includegraphics[width=\textwidth]{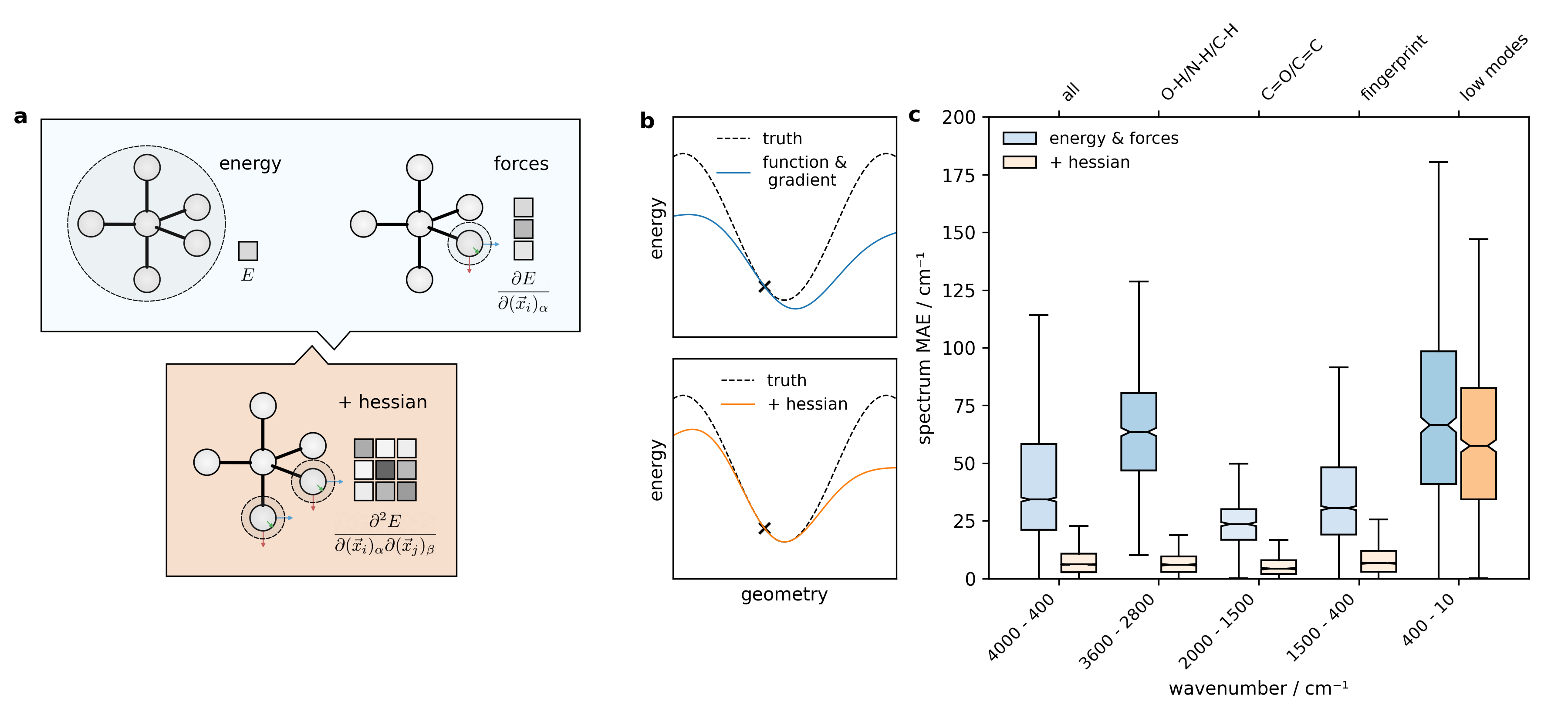}
  \caption{\textbf{a}, illustration of the global scalar/energy of a graph/molecule (left) and the vector/forces of the node/atom (right). Below is an illustration of the second-order derivative of the potential energy surface between two nodes/atoms giving a 3x3 Hessian matrix. \textbf{b}, illustrates conditioning a model to a one-dimensional function (black dashed line), where the top plot uses the function value and its gradient to train the model (filled blue line), while the bottom plot also includes the Hessian to train the model (filled orange line) demonstrating an improved approximation with higher-order derivatives. \textbf{c}, mean absolute error (MAE) between the predicted and calculated vibrational frequencies for a MLIP trained on energy and forces (blue) and a MLIP fine-tuned on Hessian data. The box-plot on the left shows the MAE for the full spectrum between $\mathrm{4000-400 \ cm^{-1}}$, subsequent box-plots represent characteristic wavenumbers, where $\mathrm{3600-2800 \ cm^{-1}}$ represents stretching vibrations for C-H, O-H and N-H bonds, $\mathrm{2000-1500 \ cm^{-1}}$ represents stretching vibrations of shorter double bonds C=C and C=O, $\mathrm{1500-400 \ cm^{-1}}$ represents the fingerprint region which contains a complex pattern of absorption bands that are specific to each molecule, $\mathrm{400-10 \ cm^{-1}}$ represents longer range molecular bending and torsional motions.}
  \label{validation}
\end{figure}

N. F. Schmitz \textit{et al.} showed that higher derivatives are more informative and aid regression models significantly (Figure \ref{validation}a and b).\cite{schmitz2022algorithmic} More recently, S. Fang \textit{et. al.} presented an E(3)-equivariant message passing graph neural network for predicting the phonon modes of the periodic crystals and molecules by evaluating the second derivative matrices of the potential energy surface.\cite{fang2023phonon} They showed that using higher order training data improved the energy models beyond the lower-order energy and force data. While the Hellmann-Feynman theorem provides an analytical method for calculating for first-order derivative of the energy eigenvalue, the second-order derivative must be calculated using finite difference methods.\cite{feynman1939forces, komornicki1993molecular} Carrying out a Taylor expansion of the potential energy surface to the second order, known as the harmonic approximation (Figure \ref{validation}a), 

\begin{equation}
    \label{eq: Taylor expansion}
    E((\vec{x}_{i})_\alpha) \approx E_0 + \sum_{(\vec{x}_{i})_\alpha,(\vec{x}_{j})_\beta} \dfrac{1}{2} \dfrac{\partial^2 E}{\partial (\vec{x}_{i})_\alpha \partial (\vec{x}_{j})_\beta} ,
\end{equation}

\noindent where $\alpha$ and $\beta$ are the Euclidean dimensions, $i$ and $j$ are atomic labels, $E$ is the energy, and $\vec{x}$ is Euclidean coordinate vector. The equation of motion for atoms close the the equilibrium geometry, governed by the dynamical matrix,

\begin{equation}
    \label{eq: dynamical matrix}
    D_{\alpha, \beta}(i,j) = \dfrac{1}{\sqrt{m_i m_j}} \sum \dfrac{\partial^2 E}{\partial (\vec{x}_{i})_\alpha \partial (\vec{x}_{j})_\beta} ,
\end{equation}

\noindent where $m$ is the atomic mass. Eigenvalue decomposition of the $(N \times 3 \times N \times 3)$ tensor, where $N$ is the number of atoms in the system, results in vectors of normal vibrational modes, and a vector of squared quantized vibrational energies,

\begin{equation}
    \label{eq: eigenvalue decomposiiton}
    \sum_{j,\beta} D_{\alpha \beta}(i,j) e_\beta^n(j) = [\omega^n]^2 e_\beta^n(j) 
\end{equation}

\noindent where $e_\beta ^n$ is the eigenvector (or eigenbasis) which represent the principal axes along which the collective motion of the atoms occurs, and $\omega$ is the vibrational energy and can be correlated to experimental infrared/Raman spectroscopy. S. Fang \textit{et. al.} also illustrated that an E(3)-equivariant message passing neural network can be fine-tuned by integrating direct second-order Hessian data or vibrational modes into the training loss function, thereby substantially improving vibrational mode property prediction for small molecules and crystals.\cite{fang2023phonon} Therefore, we have carried out numerical Hessian calculations for over 40,000 samples in each solvent environment to improve the description of the curvature of the potential energy surface at equilibrium. We validate the utility of the higher-order training data by fine-tuning an E(3)-equivariant message passing graph neural network to improve the prediction of molecular vibrational properties. We observed a refinement of stretching modes with a characteristic wavenumber $>400 \mathrm{cm^{-1}}$ after fine-tuning.

\section*{Methods}

\subsection*{DFT Calculations}

All DFT calculations were carried out using the NWChem software \cite{apra2020nwchem}. We employed the $\omega$B97X\cite {chai2008systematic} functional and the 6-31G* basis set \cite{petersson1988complete} to create data compatible with the ANI-1/ANI-1x/ANI-2x datasets. The self-consistent field (SCF) cycle was deemed converged when the changes in total energy and density were less than 10$^{-6}~\mathrm{eV}$. All molecules in the set are neutral with multiplicity equal to 1. The Mura-Knowles radial quadrature (with 49 radial points for heavy atoms, and 45 radial points for H) and Lebedev angular quadrature with 434 points for all elements were used in the integration. Structures from the QM9 dataset were optimized in vacuum and in three solvents of different polarities, tetrahydrofuran (THF) ($\epsilon_{r} = 7.6$), toluene ($\epsilon_{r} = 2.4$), and water ($\epsilon_{r} = 80.0$). The continuum solvation model based on density (SMD) was used \cite{marenich2009universal} to model the solvation effects. We used the default optimization criteria implemented in NWChem.

\subsection*{Molecule Selection}

The Hessian matrices, vibrational frequencies and modes were computed for a subset of 41,645 molecular geometries in vacuum and three solvents, applying the finite differences method, as implemented in NWChem. These candidate samples were selected from the full QM9 dataset\cite{setianto2023semi} of 133,885 using Uniform Manifold Approximation and Projection (UMAP)\cite{mcinnes2018umap} for dimensionality reduction, paired with farthest point sampling to ensure diverse configurations, as illustrated in Figure \ref{energy difference data}a (see supplementary for more details).

\subsection*{Dataset validation}

Convergence testing on the Hessian generation was performed by assessing how the frequencies change as we increase or decrease the parameters for energy, density, and gradient convergence compared to the NWChem defaults. We found that the average change in the frequency was less than 1 cm$^{-1}$. Similarly we tested the sensitivity to the finite displacement of atoms. In order to remain within the harmonic regime but avoid numerical issues in the second derivatives that may stem from too small displacements, we considered the vales of 0.005, 0.01, and 0.02 in atomic units. The average change in frequency was less than 15 cm$^{-1}$ which is below the estimated 25 cm$^{-1}$ overestimation of vibrational frequencies with the dispersion corrected version of the $\omega$B97X functional \cite{https://doi.org/10.1002/wcms.1584}, hence we opted for the middle setting of 0.01 a.u. displacement for the Hessian generation. All convergence tests were conducted on a set of 10 molecules randomly selected from the QM9 database.

\section*{Data Records}

Cheminformatics data was stored in the Hugging Face dataset format.\cite{lhoest2021datasets} For each of the four solvent environments, the data is divided into separate datasets containing vibrational analysis of 41645 optimized geometries. Table \ref{Table 1} details the data for each of the samples, where labels are associated with the QM9 molecule labelling system given by Ramakrishnan \emph{et al.}.\cite{ramakrishnan2014qm9} A link to the dataset can be found on FigShare (\url{https://figshare.com/articles/dataset/_b_Hessian_QM9_Dataset_b_/26363959}). Analysis of the datasets is displayed in Figure \ref{energy difference data}. We note that only molecules containing H, C, N, O were considered. This was because there were only 2163 molecules containing fluorine in the QM9 dataset which was not sufficient to build a good description of the chemical environment for fluorine atoms, and may have led to reduced overall precision of any models trained on our data.

\section*{Technical Validation}

The MLIP implemented in this study is structurally equivalent to the NequIP model described by Batzner et al.\cite{batzner20223nequip} (discussed in the supplementary information). Initially, the model was trained on energy and force data, achieving an average mean absolute error (MAE) of 42.44 cm$^{-1}$ for vibrational energy predictions on the validation set for energies between 4000-400 $\mathrm{cm^{-1}}$. Figure \ref{validation}c illustrates the MAE for defined energy ranges, demonstrating that the model's performance in describing motions with characteristic wavenumbers $> 400 \mathrm{cm^{-1}}$ was significantly improved by fine-tuning using higher-order gradients, resulting in an MAE of 9.49 cm$^{-1}$. While fine-tuning significantly enhanced the accuracy of vibrational energy predictions for higher energy motions, such as stretching, it did not substantially improve the predictions for longer-range bending and torsional motions. In fact, we stumbled across a fundamental limitation in eigenvalue decomposition sensitivity to perturbations in the linear transformation matrix, and showed that small eigenvalues are inherently unstable compared to larger eigenvalues (discussed in the supplementary information).

Vibrations with energies below 400 cm$^{-1}$ typically correspond to large-amplitude motions in flexible molecules or torsional motions, which are often less informative for identifying specific functional groups in organic molecules. Higher energy vibrations (above 400 cm$^{-1}$) are more likely to be associated with stretching and bending modes of bonds within molecules, which provide more distinct and useful information for chemical analysis. Therefore, we can assume that the numerically calculated low-energy vibrational modes are not only inherently sensitive, but also not particularly useful for characterizing organic molecules.

Table \ref{Table implicit solvation} illustrates the MAE for the MLIP fine-tuned with Hessian data collected using implicit solvent models as detailed in the Methods Section. The results indicate a relatively good consistency across different solvent environments. Using an implicit solvent in DFT calculations adds significant computational cost, and this dataset highlights the potential for reducing these costs by employing a MLIP fine-tuned on a subset of expensive DFT calculations. Overall, our results affirm the effectiveness of higher-order gradient methods for improving vibrational energy predictions and provide insights into the computational trade-offs involved in more advanced model fine-tuning techniques.

\section*{Code availability}

The E(3)-equivariant message passing neural network model used to predict vibrational properties is available at \url{https://github.com/google-research/e3x}. 

\bibliography{sample}

\begin{thebibliography}{10}
\urlstyle{rm}
\expandafter\ifx\csname url\endcsname\relax
  \def\url#1{\texttt{#1}}\fi
\expandafter\ifx\csname urlprefix\endcsname\relax\def\urlprefix{URL }\fi
\expandafter\ifx\csname doiprefix\endcsname\relax\def\doiprefix{DOI: }\fi
\providecommand{\bibinfo}[2]{#2}
\providecommand{\eprint}[2][]{\url{#2}}

\bibitem{PhysRev.136.B864}
\bibinfo{author}{Hohenberg, P.} \& \bibinfo{author}{Kohn, W.}
\newblock \bibinfo{journal}{\bibinfo{title}{Inhomogeneous electron gas}}.
\newblock {\emph{\JournalTitle{Phys. Rev.}}} \textbf{\bibinfo{volume}{136}}, \bibinfo{pages}{B864--B871}, \url{10.1103/PhysRev.136.B864} (\bibinfo{year}{1964}).

\bibitem{skylaris2005introducing}
\bibinfo{author}{Skylaris, C.-K.}, \bibinfo{author}{Haynes, P.~D.}, \bibinfo{author}{Mostofi, A.~A.} \& \bibinfo{author}{Payne, M.~C.}
\newblock \bibinfo{journal}{\bibinfo{title}{Introducing onetep: Linear-scaling density functional simulations on parallel computers}}.
\newblock {\emph{\JournalTitle{The Journal of chemical physics}}} \textbf{\bibinfo{volume}{122}} (\bibinfo{year}{2005}).

\bibitem{gao2020ani}
\bibinfo{author}{Gao, X.}, \bibinfo{author}{Ramezanghorbani, F.}, \bibinfo{author}{Isayev, O.}, \bibinfo{author}{Smith, J.~S.} \& \bibinfo{author}{Roitberg, A.~E.}
\newblock \bibinfo{journal}{\bibinfo{title}{Torchani: A free and open source pytorch-based deep learning implementation of the ani neural network potentials}}.
\newblock {\emph{\JournalTitle{Journal of chemical information and modeling}}} \textbf{\bibinfo{volume}{60}}, \bibinfo{pages}{3408--3415} (\bibinfo{year}{2020}).

\bibitem{smith2020ani}
\bibinfo{author}{Smith, J.~S.} \emph{et~al.}
\newblock \bibinfo{journal}{\bibinfo{title}{The ani-1ccx and ani-1x data sets, coupled-cluster and density functional theory properties for molecules}}.
\newblock {\emph{\JournalTitle{Scientific data}}} \textbf{\bibinfo{volume}{7}}, \bibinfo{pages}{134} (\bibinfo{year}{2020}).

\bibitem{devereux2020ani}
\bibinfo{author}{Devereux, C.} \emph{et~al.}
\newblock \bibinfo{journal}{\bibinfo{title}{Extending the applicability of the ani deep learning molecular potential to sulfur and halogens}}.
\newblock {\emph{\JournalTitle{Journal of Chemical Theory and Computation}}} \textbf{\bibinfo{volume}{16}}, \bibinfo{pages}{4192--4202} (\bibinfo{year}{2020}).

\bibitem{smith2017ani}
\bibinfo{author}{Smith, J.~S.}, \bibinfo{author}{Isayev, O.} \& \bibinfo{author}{Roitberg, A.~E.}
\newblock \bibinfo{journal}{\bibinfo{title}{Ani-1: an extensible neural network potential with dft accuracy at force field computational cost}}.
\newblock {\emph{\JournalTitle{Chemical science}}} \textbf{\bibinfo{volume}{8}}, \bibinfo{pages}{3192--3203} (\bibinfo{year}{2017}).

\bibitem{deng2023chgnet}
\bibinfo{author}{Deng, B.} \emph{et~al.}
\newblock \bibinfo{journal}{\bibinfo{title}{Chgnet: Pretrained universal neural network potential for charge-informed atomistic modeling}}.
\newblock {\emph{\JournalTitle{arXiv preprint arXiv:2302.14231}}}  (\bibinfo{year}{2023}).

\bibitem{batzner20223nequip}
\bibinfo{author}{Batzner, S.} \emph{et~al.}
\newblock \bibinfo{journal}{\bibinfo{title}{E (3)-equivariant graph neural networks for data-efficient and accurate interatomic potentials}}.
\newblock {\emph{\JournalTitle{Nature communications}}} \textbf{\bibinfo{volume}{13}}, \bibinfo{pages}{2453} (\bibinfo{year}{2022}).

\bibitem{musaelian2023allegro}
\bibinfo{author}{Musaelian, A.} \emph{et~al.}
\newblock \bibinfo{journal}{\bibinfo{title}{Learning local equivariant representations for large-scale atomistic dynamics}}.
\newblock {\emph{\JournalTitle{Nature Communications}}} \textbf{\bibinfo{volume}{14}}, \bibinfo{pages}{579} (\bibinfo{year}{2023}).

\bibitem{batatia2022mace}
\bibinfo{author}{Batatia, I.}, \bibinfo{author}{Kovacs, D.~P.}, \bibinfo{author}{Simm, G.}, \bibinfo{author}{Ortner, C.} \& \bibinfo{author}{Cs{\'a}nyi, G.}
\newblock \bibinfo{journal}{\bibinfo{title}{Mace: Higher order equivariant message passing neural networks for fast and accurate force fields}}.
\newblock {\emph{\JournalTitle{Advances in Neural Information Processing Systems}}} \textbf{\bibinfo{volume}{35}}, \bibinfo{pages}{11423--11436} (\bibinfo{year}{2022}).

\bibitem{gastegger2021machine}
\bibinfo{author}{Gastegger, M.}, \bibinfo{author}{Sch{\"u}tt, K.~T.} \& \bibinfo{author}{M{\"u}ller, K.-R.}
\newblock \bibinfo{journal}{\bibinfo{title}{Machine learning of solvent effects on molecular spectra and reactions}}.
\newblock {\emph{\JournalTitle{Chemical science}}} \textbf{\bibinfo{volume}{12}}, \bibinfo{pages}{11473--11483} (\bibinfo{year}{2021}).

\bibitem{reymond2015chemicalspace}
\bibinfo{author}{Reymond, J.-L.}
\newblock \bibinfo{journal}{\bibinfo{title}{The chemical space project}}.
\newblock {\emph{\JournalTitle{Accounts of Chemical Research}}} \textbf{\bibinfo{volume}{48}}, \bibinfo{pages}{722--730} (\bibinfo{year}{2015}).

\bibitem{ramakrishnan2014qm9}
\bibinfo{author}{Ramakrishnan, R.}, \bibinfo{author}{Dral, P.~O.}, \bibinfo{author}{Rupp, M.} \& \bibinfo{author}{Von~Lilienfeld, O.~A.}
\newblock \bibinfo{journal}{\bibinfo{title}{Quantum chemistry structures and properties of 134 kilo molecules}}.
\newblock {\emph{\JournalTitle{Scientific data}}} \textbf{\bibinfo{volume}{1}}, \bibinfo{pages}{1--7} (\bibinfo{year}{2014}).

\bibitem{ward2021graph}
\bibinfo{author}{Ward, L.} \emph{et~al.}
\newblock \bibinfo{journal}{\bibinfo{title}{Graph-based approaches for predicting solvation energy in multiple solvents: open datasets and machine learning models}}.
\newblock {\emph{\JournalTitle{The Journal of Physical Chemistry A}}} \textbf{\bibinfo{volume}{125}}, \bibinfo{pages}{5990--5998} (\bibinfo{year}{2021}).

\bibitem{schmitz2022algorithmic}
\bibinfo{author}{Schmitz, N.~F.}, \bibinfo{author}{Müller, K.-R.} \& \bibinfo{author}{Chmiela, S.}
\newblock \bibinfo{journal}{\bibinfo{title}{Algorithmic differentiation for automated modeling of machine learned force fields}}.
\newblock {\emph{\JournalTitle{The Journal of Physical Chemistry Letters}}} \textbf{\bibinfo{volume}{13}}, \bibinfo{pages}{10183--10189} (\bibinfo{year}{2022}).

\bibitem{fang2023phonon}
\bibinfo{author}{Fang, S.}, \bibinfo{author}{Geiger, M.}, \bibinfo{author}{Checkelsky, J.} \& \bibinfo{author}{Smidt, T.}
\newblock \bibinfo{title}{Phonon predictions with e(3)-equivariant graph neural networks}.
\newblock In \emph{\bibinfo{booktitle}{AI for Accelerated Materials Design - NeurIPS 2023 Workshop}} (\bibinfo{year}{2023}).

\bibitem{feynman1939forces}
\bibinfo{author}{Feynman, R.~P.}
\newblock \bibinfo{journal}{\bibinfo{title}{Forces in molecules}}.
\newblock {\emph{\JournalTitle{Physical review}}} \textbf{\bibinfo{volume}{56}}, \bibinfo{pages}{340} (\bibinfo{year}{1939}).

\bibitem{komornicki1993molecular}
\bibinfo{author}{Komornicki, A.} \& \bibinfo{author}{Fitzgerald, G.}
\newblock \bibinfo{journal}{\bibinfo{title}{Molecular gradients and hessians implemented in density functional theory}}.
\newblock {\emph{\JournalTitle{The Journal of chemical physics}}} \textbf{\bibinfo{volume}{98}}, \bibinfo{pages}{1398--1421} (\bibinfo{year}{1993}).

\bibitem{apra2020nwchem}
\bibinfo{author}{Apra, E.} \emph{et~al.}
\newblock \bibinfo{journal}{\bibinfo{title}{Nwchem: Past, present, and future}}.
\newblock {\emph{\JournalTitle{The Journal of chemical physics}}} \textbf{\bibinfo{volume}{152}} (\bibinfo{year}{2020}).

\bibitem{chai2008systematic}
\bibinfo{author}{Chai, J.-D.} \& \bibinfo{author}{Head-Gordon, M.}
\newblock \bibinfo{journal}{\bibinfo{title}{Systematic optimization of long-range corrected hybrid density functionals}}.
\newblock {\emph{\JournalTitle{The Journal of chemical physics}}} \textbf{\bibinfo{volume}{128}} (\bibinfo{year}{2008}).

\bibitem{petersson1988complete}
\bibinfo{author}{Petersson, a.} \emph{et~al.}
\newblock \bibinfo{journal}{\bibinfo{title}{A complete basis set model chemistry. i. the total energies of closed-shell atoms and hydrides of the first-row elements}}.
\newblock {\emph{\JournalTitle{The Journal of chemical physics}}} \textbf{\bibinfo{volume}{89}}, \bibinfo{pages}{2193--2218} (\bibinfo{year}{1988}).

\bibitem{marenich2009universal}
\bibinfo{author}{Marenich, A.~V.}, \bibinfo{author}{Cramer, C.~J.} \& \bibinfo{author}{Truhlar, D.~G.}
\newblock \bibinfo{journal}{\bibinfo{title}{Universal solvation model based on solute electron density and on a continuum model of the solvent defined by the bulk dielectric constant and atomic surface tensions}}.
\newblock {\emph{\JournalTitle{The Journal of Physical Chemistry B}}} \textbf{\bibinfo{volume}{113}}, \bibinfo{pages}{6378--6396} (\bibinfo{year}{2009}).

\bibitem{setianto2023semi}
\bibinfo{author}{Setianto, S.}, \bibinfo{author}{Panatarani, C.}, \bibinfo{author}{Singh, D.} \& \bibinfo{author}{Joni, I.~M.}
\newblock \bibinfo{journal}{\bibinfo{title}{Semi-empirical infrared spectra simulation of pyrene-like molecules insight for simple analysis of functionalization graphene quantum dots}}.
\newblock {\emph{\JournalTitle{Scientific Reports}}} \textbf{\bibinfo{volume}{13}}, \bibinfo{pages}{2282} (\bibinfo{year}{2023}).

\bibitem{mcinnes2018umap}
\bibinfo{author}{McInnes, L.}, \bibinfo{author}{Healy, J.} \& \bibinfo{author}{Melville, J.}
\newblock \bibinfo{journal}{\bibinfo{title}{Umap: Uniform manifold approximation and projection for dimension reduction}}.
\newblock {\emph{\JournalTitle{arXiv preprint arXiv:1802.03426}}}  (\bibinfo{year}{2018}).

\bibitem{https://doi.org/10.1002/wcms.1584}
\bibinfo{author}{Zapata~Trujillo, J.~C.} \& \bibinfo{author}{McKemmish, L.~K.}
\newblock \bibinfo{journal}{\bibinfo{title}{Meta-analysis of uniform scaling factors for harmonic frequency calculations}}.
\newblock {\emph{\JournalTitle{WIREs Computational Molecular Science}}} \textbf{\bibinfo{volume}{12}}, \bibinfo{pages}{e1584}, \url{https://doi.org/10.1002/wcms.1584} (\bibinfo{year}{2022}).
\newblock \eprint{https://wires.onlinelibrary.wiley.com/doi/pdf/10.1002/wcms.1584}.

\bibitem{lhoest2021datasets}
\bibinfo{author}{Lhoest, Q.} \emph{et~al.}
\newblock \bibinfo{journal}{\bibinfo{title}{Datasets: A community library for natural language processing}}.
\newblock {\emph{\JournalTitle{arXiv preprint arXiv:2109.02846}}}  (\bibinfo{year}{2021}).

\end{thebibliography}

\section*{Acknowledgements} 

This work was supported by the Hartree National Centre for Digital Innovation, a collaboration between STFC and IBM.

\section*{Author contributions statement}

NJW carried out sample selection from the QM9 dataset, as well as creating, training and validating the E(3)-equivariant graph neural network MLIP. VZ supervised the high-throughput generation of DFT data. NJW, LK, LS, and VZ generated the data. LK and LS performed convergence testing and data validation. EPK helped to conceive the project, aided with algorithms for sample selection. NJW, LK, LS, and VZ contributed equally to preparing the manuscript. 

\section*{Competing interests}

No competing interests.  

\section*{Figures \& Tables}

\begin{table}[h!]
\centering
\begin{tabular}{||c c c c||} 
 \hline
 Label & Description & Units & Typing \\ [0.5ex] 
 \hline\hline
 label & QM9 database label & - & string \\ 
 atomic\_numbers & Atomic numbers ($N\times 1$)& - & list \\
 positions & Atomic positions ($N\times 3$) & \r{A} & array \\
 energy & Potential energy & $\mathrm{eV}$ & float \\
 forces & Forces ($N\times 3$) & $\mathrm{eV}$/\r{A} & array \\
 Hessian & Hessian matrix ($N\times 3\times N\times 3$) & $\mathrm{eV}$/\r{A}$^2$ & array \\
 frequencies & Vibrational frequencies ($3N\times 1$) & $\mathrm{cm^{-1}}$ & array \\
 normal\_modes & Normal vibrational modes ($3N\times 3N$) & - & array \\ [1ex] 
 \hline
\end{tabular}
\caption{Table with labels, descriptions, units, and typing information for the Hessian QM9 dataset.}
\label{Table 1}
\end{table}

\begin{table}[h!]
\centering
\begin{tabular}{||c c c c c||} 
 \hline
 Solvent & E / meV & F / meV/\r{A} & H / meV/\r{A}$^2$ & $\omega$ (4000-400) / $\mathrm{cm^{-1}}$ \\ [0.5ex] 
 \hline\hline
 vacuum & 36 & 35 & 70 & 9.49 \\
 thf & 29 & 28 & 67 & 8.72 \\
 toluene & 32 & 29 & 62 & 8.30 \\\
 water & 35 & 28 & 54 & 8.12 \\  
 \hline
\end{tabular}
\caption{Mean absolute error for energy, forces, Hessian, and vibrational energy (4000-400 $\mathrm{cm^{-1}}$) on the validation set for the \texttt{e3x} model fine-tuned with Hessian data in different implicit solvents.}
\label{Table implicit solvation}
\end{table}

\begin{figure}[ht]
  \centering
  \includegraphics[width=\textwidth]{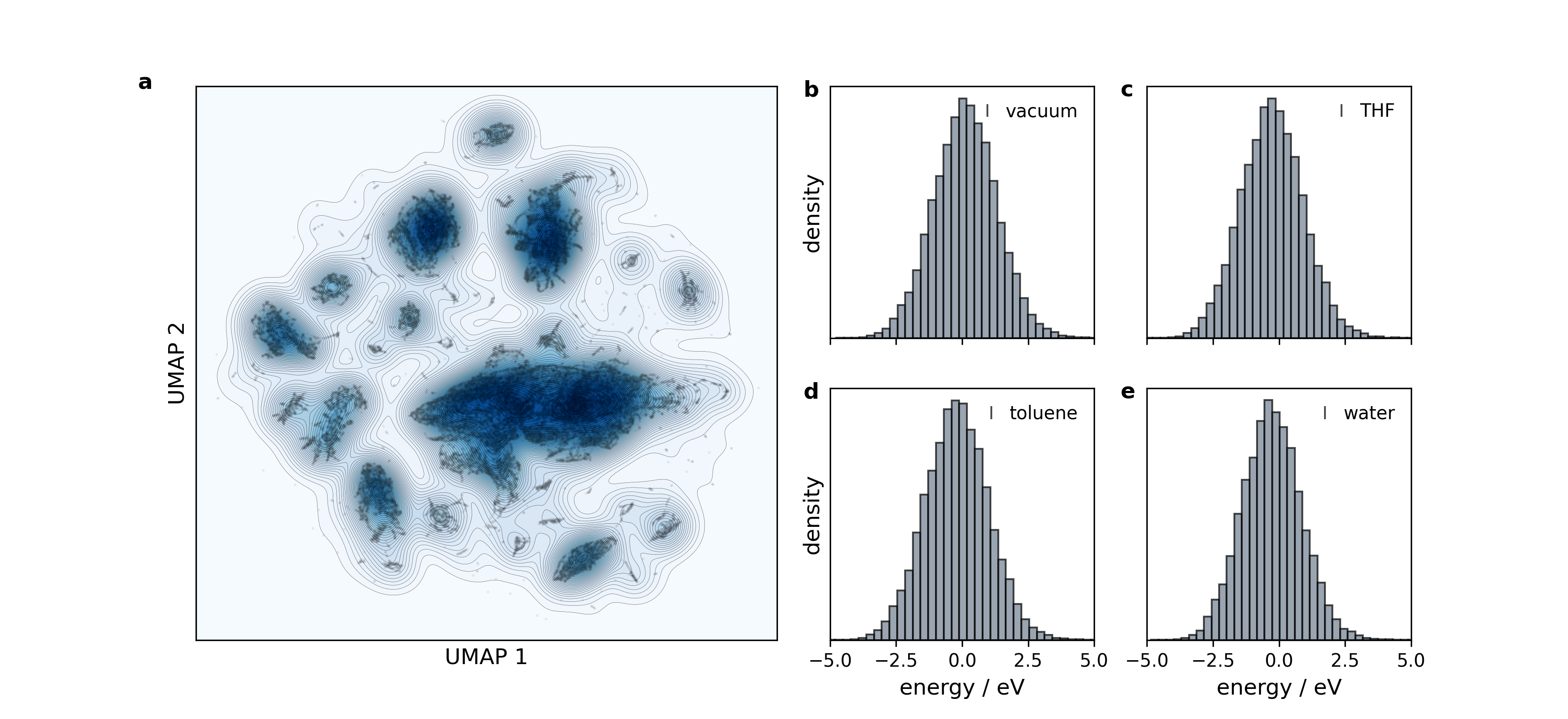}
  \caption{\textbf{a}, UMAP density plot of the QM9 dataset using SOAP descriptors where contour lines show the density of the whole dataset and the black dots show the 40,000 sampled data-points using farthest point sampling. \textbf{b}-\textbf{e}, histograms of the energy differences, eV, from linear regression of atomic energies for sampled ground state configurations in different environments. The energy differences are calculated as deviations from a reference model where the atomic energies of hydrogen, carbon, nitrogen, and oxygen are approximately $-16.67$, $-1035.81$, $-1489.65$, and $-2047.05$ eV, respectively. In vacuum, \textbf{b}, the mean energy difference of $-0.05$ eV and a standard deviation of 1.19 eV, in THF, \textbf{c}, the mean energy difference is $-0.44$ eV with a standard 1.17 eV, in toluene, \textbf{d}, the mean is $-0.37$ eV and the standard deviation is 1.19 eV,  and in water, \textbf{e}, the mean energy difference is $-0.45$ eV and the standard deviation is 1.11 eV. 
  }
  \label{energy difference data}
\end{figure}

\end{document}